\documentclass[]{spie}  

\usepackage{amsmath,amsfonts,amssymb}
\usepackage{graphicx}
\usepackage[colorlinks=true, allcolors=blue]{hyperref}
\usepackage{hyperref}
\usepackage{dsfont}
\usepackage{tikz}
\usetikzlibrary{shapes,arrows}
\usepackage{comment}
\newcommand{\rmd}{\textrm{d}}

\title{GPI 2.0 : Optimizing reconstructor performance in simulations and preliminary contrast estimates}

\author[a]{Alexander Madurowicz}
\author[a]{Bruce Macintosh}
\author[b]{Lisa Poyneer}
\author[d]{Duan Li}
\author[a]{Jean-Baptiste Ruffio}
\author[c]{Jean-Pierre Veran}
\author[b]{S. Mark Ammons}
\author[d]{Dmitry Savransky}
\author[e]{Jeffrey Chilcote}
\author[f]{Jerome Maire}
\author[f]{Quinn Konopacky}
\author[a,g]{Robert J. De Rosa}
\author[c]{Christian Marois}
\author[h]{Marshall Perrin}
\author[h]{Laurent Pueyo}

\affil[a]{Kavli Institute for Particle Astrophysics and Cosmology, Stanford University, Stanford, CA 94305, USA}
\affil[b]{Lawrence Livermore National Laboratory, Livermore, CA 94551, USA}
\affil[c]{National Research Council of Canada Herzberg, 5071 West Saanich Rd, Victoria, BC, V9E 2E7, Canada}
\affil[d]{Sibley School of Mechanical and Aerospace Engineering, Cornell University, Ithaca, NY 14853, USA}
\affil[e]{Department of Physics, University of Notre Dame, 225 Nieuwland Science Hall, Notre Dame, IN, 46556, USA}
\affil[f]{Center for Astrophysics and Space Science, University of California San Diego, La Jolla, CA 92093, USA}
\affil[g]{European Southern Observatory, Alonso de Cordova 3107, Vitacura, Santiago, Chile}
\affil[h]{Space Telescope Science Institute, Baltimore, MD 21218, USA}

\authorinfo{Send correspondence to Alexander Madurowicz amaduro@stanford.edu}
\begin{document} 
\maketitle

\begin{abstract}
During its move from the mountaintop of Cerro Pachon in Chile to the peak of Mauna Kea in Hawaii, the Gemini Planet Imager will make a pit stop to receive various upgrades, including a pyramid wavefront sensor. As a highly non-linear sensor, a standard approach to linearize the response of the pyramid is induce a rapid circular modulation of the beam around the pyramid tip, trading off sensitivity for robustness during high turbulence. Using high temporal resolution Fourier Optics based simulations, we investigate phase reconstruction approaches that attempt to optimize the performance of the sensor with a dynamically adjustable modulation parameter. We have studied the linearity and gain stability of the sensor under different modulation and seeing conditions, and the ability of the sensor to correct non-common-path errors. We will also show performance estimates which includes a comparative analysis of the atmospheric columns above the two mountains, as well as the Error Transfer Functions of the two systems.
\end{abstract}

\section{Introduction}

The Gemini Planet Imager (GPI) is an instrument capable of directly imaging and spectroscopically characterizing young, massive extrasolar planets. \cite{Macintosh2014} GPI operated for roughly five years on the Gemini South telescope, and observed a large survey of the most accessible young nearby systems \cite{nielsen2019} to constrain giant planet demographics and formation mechanisms. In part due to the availability of targets and changing observatory priorities, GPI is planned to move from Gemini South in Chile to Gemini North on Mauna Kea, in Hawaii. \cite{Chilcote2018} During the transition, a number of upgrades to the instrument are planned to boost its performance, including a pyramid wavefront sensor (WFS), faster real time computer (RTC), zero noise EMCCDs, a low spectral resolution broadband filter mode, and modern redesigned apodized pupil Lyot coronagraphs, among other changes. In this paper, we will focus on the upgrades of the pyramid wavefront sensor and real time computer. Section 2 broadly covers the pyramid wavefront sensor, developing a Fourier-Optics based model of the instrument and investigating phase reconstruction approaches in simulations, as well as sections on optical gain calibration and non-common-path aberration correction. Section 3 investigates improvements to the Error Transfer Function on the system as a whole due to improvements in compute delay from the faster RTC, and the effect this could have on the final performance of the instrument using a comparative analysis of the atmospheres on the two sites.

\section{Phase Reconstruction with a Pyramid WFS}

The pyramid wavefront sensor is a well-known \cite{VERINAUD2004,chambouleyron2020,deo2018,Korkiakoski08,Hutterer_2019,Shatokhina2020} substitute to the classical Shack-Hartmann sensor, with notable performance improvements \cite{correia2020} which can be attributed to favorable error propagation properties. \cite{Plantet2015} While the pyramid sensor is known to have a non-linear response to high amplitude phase aberrations, \cite{Landman2020} many AO system operate in a closed-loop fashion \cite{esposito2000} \cite{ragazzoni1999} which boosts the pyramid efficacy as the system approaches the diffraction limit. Additionally, a common tactic to further linearize the pyramid sensor is to induce a rapid modulation of the beam around the pyramid tip \cite{burvall2007}, which trades off sensitivity for linear dynamic range, which can potentially even be tuned during operation \cite{akondi2013} to respond to a dynamically changing atmosphere.

\subsection{Description of the Optical Problem}

In order to model the telescope and AO system response to an evolving atmosphere, we construct a Fourier Optics based approach to modeling the Pyramid Wavefront Sensor. This approach considers three optical planes, the pupil plane $(x,y)$, the Pyramid Optic Plane $(\xi,\eta)$, and the WFS image plane $(\alpha,\beta)$. The complex electric field entering the pupil $U_\textrm{atm}(x,y)$ is a function of the atmospheric model described in detail in references \cite{madurowicz2018,madurowicz2019} and calibrated in Appendix A. This model describes Fresnel Propagation of light through frozen-flow layered Kolmogorov phases screens with variable wind velocities, calibrated to produce a desired value of the Fried Parameter. 

In addition, a rapid modulation of the beam around the tip of the pyramid is achieved by means of an ideal tip-tilt phase mirror $\phi_\textrm{mod}$ conjugated to the pupil plane, and the transmission function of the pupil $T_\textrm{pupil}$ is an idealized 8-meter diameter circle with a secondary obscuration corresponding to the Gemini Telescope secondary mirror. The differential piston effect of the secondary supporting spiders \cite{hutterer2018} for the Gemini pupil is not resolvable at the current resolution of our simulation, due to their very thin profile they are are smaller than a single pixel is wide. There is additionally another phase term $\phi_\textrm{DM}$ that will be controlled by the system deformable mirror to mitigate the atmospheric phase. In total, the complex electric field at the end of the first plane is
\begin{equation}
    U_\textrm{pupil} = U_\textrm{atm}T_\textrm{pupil}e^{i\phi_\textrm{mod}}e^{i\phi_\textrm{DM}}
\end{equation}
where
\begin{equation}
    T_\textrm{pupil} = 
    \begin{cases}
    1, \textrm{if} \, R_\textrm{sec} \leq \sqrt{x^2 + y^2} \leq R_\textrm{tel} \\
    0, \textrm{otherwise}
    \end{cases}
\end{equation}
and
\begin{equation}
    \phi_\textrm{mod} = \alpha_\textrm{mod}\frac{2\pi}{D_\textrm{tel}}(x\cos{\psi}+y\sin{\psi}).
\end{equation}
Here $\alpha_\textrm{mod}$ is the modulation radius in units of $\lambda/D_\textrm{tel}$ and $\psi \in [0,2\pi)$ is the modulation azimuthal parameter, which represents the spot traveling around the circle on the pyramid tip. This is implicitly making the assumption that $U_\textrm{atm}(x,y)$ is not a function of $\psi$, or that the atmosphere is frozen in place during the sub-modulation timesteps. This assumption is justified by noting that the atmospheric timescale $\tau_0$ in the worst case is at least a few milliseconds \cite{TMTsite} and for an AO system running at 1 kHz, an entire modulation cycle happens faster than a millisecond. In principle it is possible to resolve the temporal error that this assumption introduces but that drastically increases the computation necessary for the atmospheric model.

The Pyramid Optic itself is modeled as a phase mask, which is not strictly true. This is equivalent to assuming that the pyramid optic's physical height is much smaller than the focal length of the beam, and a more robust treatment would introduce slight defocus as the beam approaches the edge of the pyramid. However, this approach to model pyramid is used rather extensively, \cite{Korkiakoski08,Hutterer_2019,Shatokhina2020}  and is quite successful. Using the knowledge that the image plane electric field distribution is given by the inverse Fourier Transform of the complex illumination of the pupil \cite{madurowicz2019}, we can model the final wavefront sensor intensity distribution with the following quintuple integral
\begin{equation}
    I_\textrm{WFS}(\alpha,\beta) = \frac{1}{2\pi}\int_0^{2\pi}\rmd\psi \Big|\mathcal{F}^{-1}\big[\mathcal{F}^{-1}\big[U_\textrm{pupil}\big]e^{i\phi_\textrm{PYWFS}}\big]\Big|^2,
\end{equation}
which averages the instantaneous intensity during each modulation azimuthal angle $\psi$ during the observation. Here $\mathcal{F}^{-1}$ is the inverse Fourier Transform given by:
\begin{equation}
\mathcal{F}^{-1}\big[f(x,y)\big](\xi,\eta) = \iint_{-\infty}^{\infty}\rmd x \rmd y f(x,y) e^{i(x\xi+y\eta)},
\end{equation}
and $\phi_\textrm{PYWFS}$ is the pyramid phase mask, given by:
\begin{equation}
    \phi_\textrm{PYWFS} = \alpha_\textrm{PY}
    \begin{cases}
    |\xi + \eta|, \textrm{quadrants 1 and 3} \\
    |\xi - \eta|, \textrm{quadrants 2 and 4},
    \end{cases}
\end{equation}
with $\alpha_\textrm{PY}$ a particular constant which describes the opening angle of the pyramid, and the quadrants in question are the standard, with quadrant 1 satisfying $\xi > 0$ and $\eta > 0$, increasing counterclockwise. For $\alpha_\textrm{PY} = 0$, the four resulting pupil re-images will be superimposed at the origin, as if no pyramid exists, but with $\alpha_\textrm{PY} > 0$, the four pupil image begin to separate, with large values driving the reimaged pupils to wider angles in the $(\alpha,\beta)$ plane. 

For our particular simulation, the value of $\alpha_\textrm{PY}$ we use is naturally defined by the simulation box pixel's resolution. For a simulation box with $(N,N)$ pixels, with half-dimension $D$ meters, such that a single pixel occupies $\rmd \xi = 2D/N$ meters, the maximum angular scale in the Fourier plane is $\lambda/(2\rmd \xi)$ for monochromatic light at wavelength $\lambda$. If we would like the four re-imaged pupils to be nested equidistantly from each other, such that each is in the center of its own sub-quadrant of the final WFS image plane, they must be located at $\lambda/(4\rmd \xi)$ along each axis, since there are $2\pi/\lambda$ radians per wavelength,
\begin{equation}
    \alpha_\textrm{PY} = \frac{\pi}{2\rmd \xi}.
\end{equation}
This approach essentially treats the pyramid optical element as being the same physical dimension as the telescope pupil, but does not  affect the result. Additionally, due the nature of periodicity due to the Fourier Transform, there are an infinite number of solutions for $\alpha_\textrm{PY}$ which produce equivalent answers, as higher values for the slope will push the re-imaged pupils outside of the box dimension and back into the simulation on the other side, but some of these values cause the relative locations of the pupils to be flipped along both axes, and so the reconstruction may need to be mirrored along both axes to remain consistent.

In principle, equation (4) is not (complex) analytic, because of discontinuities and the complex conjugate operation, but it may still be possible to analyze further on analytic subdomains using a Fourier decomposition of $U_\textrm{atm}(x,y)$. Additionally, equation (4) is non-linear. Even though both the integral operator, the inverse Fourier Transform operator, and multiplication by a complex number $e^{i\gamma}$ are linear, there are multiple non-linear components including the absolute magnitude $|\cdot|^2$, the discontinuities at the edge of the pupil and the pyramid faces, and the complex exponentiation of the phase $e^{i\phi}$ itself. 

However, the standard approach in AO systems is to assume that the sensor operator is a linear function, and to measure the forward operator with a matrix and then invert it by means of the Singular Value Decomposition or some other regularized inverse. \cite{vanDam2004} This approach relies on an assumption that the value of the phase entering the pupil is small, so one can think of a kind of Taylor expansion around zero phase, where small perturbations are well modeled by the first term in the expansion. This is only possible to achieve during closed-loop operation, where the vast majority of the phase has already been mitigated by the system deformable mirror. So, incidentally, in order to solve an impossibly hard problem, one must first assume that one already has, and it suddenly becomes possible.

\begin{figure}[h!]
    \centering
    \includegraphics[width=\textwidth]{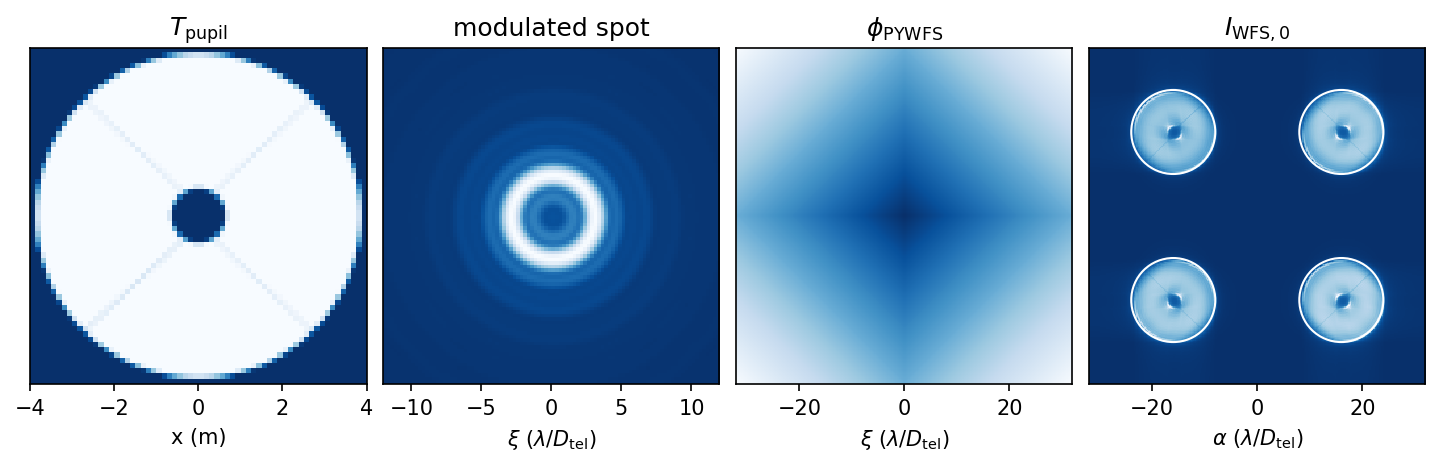}
    \caption{A demonstration of measuring the WFS zero phase state in our simulation. One can see that the very narrow Gemini pupil spiders in our low resolution simulation act to reduce $T_\textrm{pupil} < 1$ in some pixels which they overlap, but are not wide enough to be completely opaque pixels. The modulated spot is depicting the average intensity in the pyramid optic plane $(\xi,\eta)$ averaged over the modulation $\psi$, although the full complex electric field is used when adding $\phi_\textrm{PYWFS}$ and computing the final WFS intensity. One can see the zero point reference state has the four re-imaged pupils, each centered in its own quadrant. However, the Intensity distribution is not flat, and includes bright and dark regions due to diffraction off the spiders and edge of the pupil. The four re-imaged pupil are circled in white to highlight the region of interest.}
    \label{fig:optical_path}
\end{figure}

This zero phase condition $\phi_\textrm{atm} + \phi_\textrm{DM} = 0$ implies that $U_\textrm{pupil,0} = T_\textrm{pupil}e^{i\phi_\textrm{mod}}$ if we ignore the amplitude aberrations produced by Fresnel propagation in the atmosphere, which is the simplest possible case to evaluate the WFS integral in equation (4). The resulting PSF in the pyramid optic plane $\mathcal{F}^{-1}[U_\textrm{pupil,0}]$ will be a shifted radial sinc function, or a tilted airy disk if we ignore the secondary obscuration. The secondary obscuration slightly modifies the shape of the PSF by introducing oscillations on alternating airy rings. The final WFS intensity distribution $I_\textrm{WFS,0}$ in this case will be important to measure, as it will serve as the reference zero-point during closed loop operation. This is demonstrated in Figure \ref{fig:optical_path}.

\subsection{Modal Basis Interaction Matrix}

Once the zero point state of the sensor has been measured, the next step to formulating a reconstruction process is to measure an interaction matrix for small phase perturbations on the DM. To do this, a set of basis vectors for the DM must be chosen. There are many possible basis sets, including actuator pokes, sines and cosines in the Fourier basis, the set of Zernike Polynomials, but since many AO systems are dominated by time-lag error, it would be nice if the basis set $\textit{efficiently}$ represented the phase distributions we would like to reproduce. An efficient representation minimizes the number of basis elements needed to reach a particular level of accuracy. This is achievable through the use of principal components. \cite{Jolliffe2016}

\begin{figure}[h!]
    \centering
    \includegraphics[width=.8\textwidth]{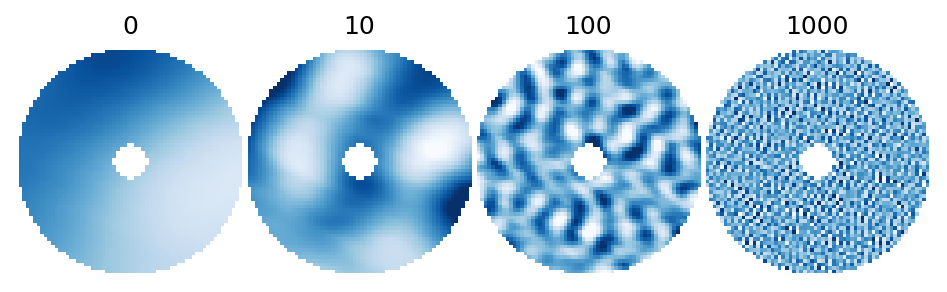}
    \caption{Some examples of principle components used to form an orthonormal basis over the DM vector space. The number in the title represents the modal index, ordered by decreasing eigenvalue.}
    \label{fig:pca}
\end{figure}

By treating the simulated atmosphere as a dataset of realizations in the vector space spanned by the DM actuators, and computing the covariance matrix of this data, one can find the principle components by computing the eigenvectors of the covariance matrix. The DM vector space must include every pixel where the pupil transmission function is non-zero, and so the number of modes or the dimension of the vector space is
\begin{equation}
N_\textrm{modes} = \sum_\textrm{pixels} \Big[ T_\textrm{pupil} > 0 \Big].
\end{equation}
If the simulation runs for a length of $L$ timesteps, then the atmospheric data $\mathbf{A}$ can be represented by a matrix of dimension $(L,N_\textrm{modes})$, as there are L realizations each of size $N_\textrm{modes}$. Then, the covariance matrix $\mathbf{C}_{mn}$ is an $(N_\textrm{modes},N_\textrm{modes})$ matrix, and can be computed directly from the data with
\begin{equation}
\mathbf{C}_{mn} = \frac{1}{L-1} \sum_{l=1}^{L} (\mathbf{A}_{lm} - \Bar{\mathbf{A}}_m)(\mathbf{A}_{ln} - \Bar{\mathbf{A}}_n).
\end{equation}
Here $\mathbf{A}_{ln}$ is the $l^{th}$ realization of the $n^{th}$ mode, and $\Bar{\mathbf{A}}_n$ is the average over all $L$ realizations of the $n^{th}$ mode, which is computable with
\begin{equation}
    \Bar{\mathbf{A}}_n = \frac{1}{L} \sum_{l=1}^{L} A_{ln}.
\end{equation}
With the covariance matrix in hand, the principle components are just its eigenvectors, which are the solutions $\mathbf{v}$ to the equation
\begin{equation}
    \mathbf{C}\mathbf{v}=\gamma\mathbf{v},
\end{equation}
which, when sorted by corresponding eigenvalue $\gamma$, is an orthonormal basis where each successive basis element explains the maximum remaining variance left in the data. Examples of a few different principle components are demonstrated in Figure \ref{fig:pca}

By combining all of the eigenvectors $\mathbf{v}$ into a matrix , we can construct an object which maps a vector of modal coefficients into the DM vector space
\begin{equation}
    \mathbf{V} = 
    \begin{bmatrix}
    \mathbf{v}_1 \\
    \mathbf{v}_2 \\
    \vdots \\
    \mathbf{v}_{N_\textrm{modes}} \\
    \end{bmatrix}.
\end{equation}
This matrix $\mathbf{V}$ can be multiplied onto one-hot encoding vectors $\mathds{1}_j = [0, 0, ..., 1, ..., 0]$ to immediately recover the $j^{th}$ principal component and represents the DM vector basis. In general, there will be some DM state given by a list of modal coefficients $\textbf{v}_{\textrm{DM}}$, and the DM phase will be produced using $\phi_\textrm{DM} = \mathbf{V}\mathbf{v}_\textrm{DM}$. Because the principal components process ensures the orthonormality of the basis vectors $\mathbf{v}_i\cdot\mathbf{v}_j = \delta_{ij}$ at least to numerical precision, this basis is quite excellent. In principle it is possible to use an non-orthogonal basis set (this is actually an oxymoron, it is really just a spanning set) to describe DM space, but this causes serious issues with the interaction matrix framework. Having linearly independent basis vectors is critical to be able to calibrate the sensor response function to perturbations, as a single mode on the DM does not "mix" with other modes.

To compute the interaction matrix $\mathbf{I}$, it is as straightforward as simulating each mode on the DM, when the atmospheric aberrations are removed. As long as the phase perturbations are small, the sensor operates in the linear regime and is well modeled by the interaction matrix. It is not necessary to extract the four re-imaged pupils from the WFS plane or to compute the sensor slopes by adding and subtracting the relevant quadrants in this map, but in practice this reduces the rectangularity of the interaction matrix, as the number of pixels inside the four re-imaged pupils is already four times greater than the relevant number of modes being controlled. This extraction is achieved by means of a boolean operation on the WFS intensity, corresponding to four copies of the pupil boolean map which have been truncated and aligned with the intensity distribution in a calibration step known as registration. For a real pyramid optic, slight deviations in the slope of the faces may cause the registration to be inexact, yet in our simulation we can guarantee the location of the re-imaged pupils to be centered in each sub-quadrant of the image.

\begin{figure}[h!]
    \centering
    \includegraphics[width=\textwidth]{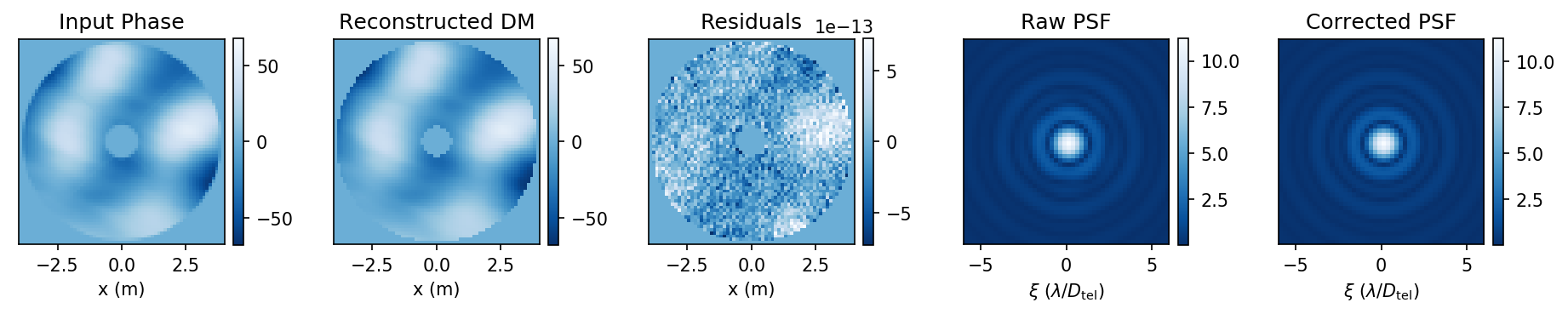}
    \includegraphics[width=\textwidth]{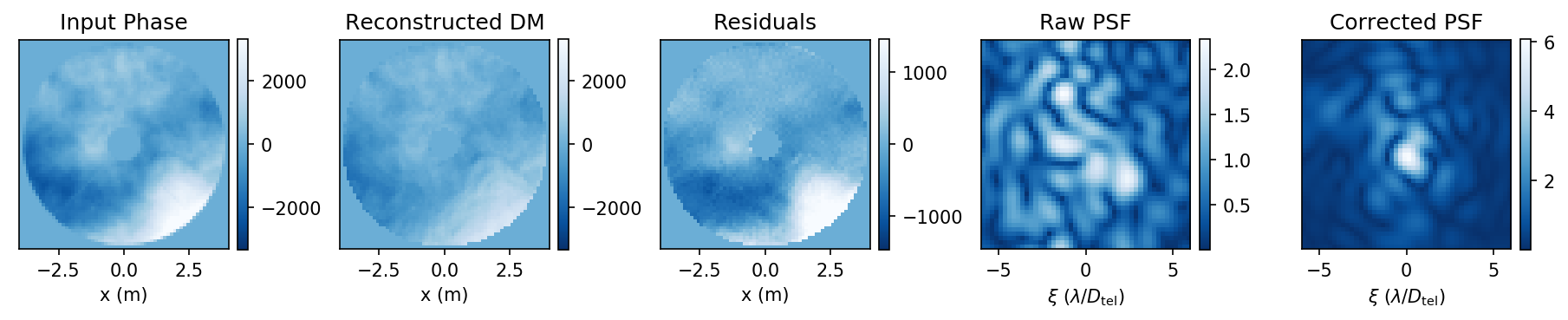}
    \caption{Testing the reconstruction process on a small modal perturbation and a fully aberrated atmosphere. Column 1 contains the phase placed on the DM for testing, while column 2 contains the reconstructed phase using the inverse of the interaction matrix acting on the extracted WFS gradients, and column 3 is the difference between the two. All colormaps for columns 1-3 are in units of nanometers. Columns 4 and 5 contain the original PSF produced by the aberration compared to PSF when the reconstruction is subtracted, using the residual phase error map. The colormaps for the PSFs are proportional to the square root of the intensity. Since the simulation contains no WFS noise the reconstruction is nearly perfect for the small perturbation, with residuals of order $10^{-13}$ nm, but does not work as well for the fully aberrated case. While it corrects the shape of the DM, the reconstructor does not properly estimate the magnitude of the aberration, and the residuals are still on the order of $\sim 50 \%$ of the input. This is due to the non-linearity of large phase perturbations, but as the loop closes over multiple reconstruction steps, will still drive the residual phase towards zero. }
    \label{fig:recon}
\end{figure}

After the WFS intensity pixels are extracted, and the slopes are computed, the gradients are normalized by the average intensity across all of the re-imaged pupils, in order to provide a brightness correction for different stars. The gradients are then referenced to gradients computed on the flat WFS state $I_\textrm{WFS,0}$, and stacked into a matrix, which is inverted by means of the singular value decomposition. The interaction matrix can be decomposed into $\mathbf{I} = U\Sigma V^*$, which implies the existence of the pseudo-inverse $\mathbf{I}^+ = V\Sigma^{-1}U^*$. Here $U$ and $V$ are square, real, orthonormal, and unitary matrices, and $\Sigma$ is a diagonal matrix containing the singular values. This is a useful method to decompose the linear transformation $\mathbf{I}$, because $U$ and $V$ are unitary operators, they can be thought of as acting to rotate the basis elements of the space, while $\Sigma$ acts to stretch the rotated vector along the intermediary axis. This combination of rotate, stretch, derotate naturally allows one to find the pseudo-inverse, by the means of de-rotating, un-stretching, and re-rotating. A couple of example tests using the pseudo-inverse of the interaction matrix are shown in Figure \ref{fig:recon}.

\subsection{Discussion on Optical Gain calibration}

As demonstrated in the previous section, when the phase aberrations in the pupil are large, such as when the uncorrected atmosphere is present, the reconstructor based on small phase perturbations does not accurately reproduce the inputs. This behavior due to the non-linearity of the sensor has been previously called the optical gain problem. \cite{chambouleyron2020,deo2018}  

\begin{figure}[h!]
    \centering
    \includegraphics[width=\textwidth]{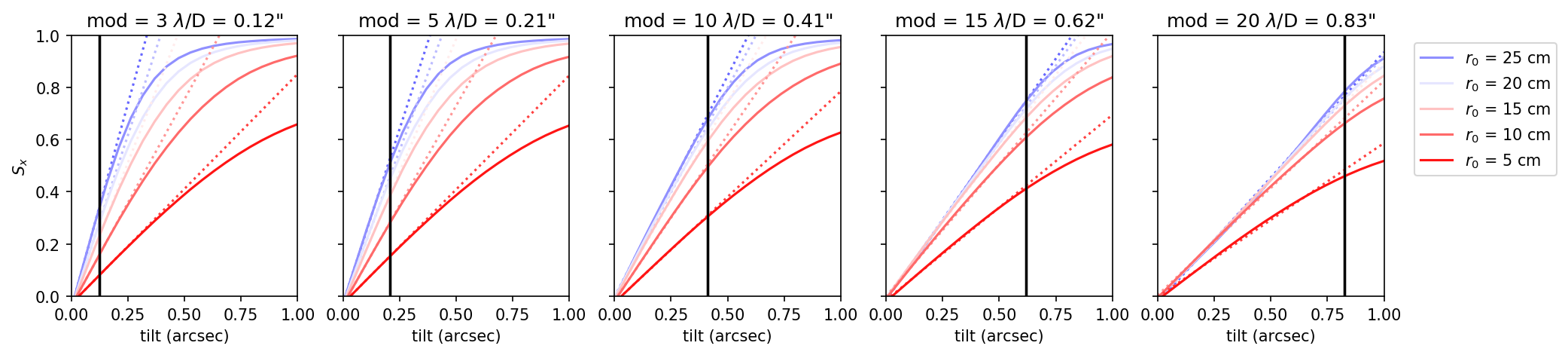}
    \caption{Pyramid Wavefront sensor response curves to injected tilt on an aberrated PSF. Each plot corresponds to a different modulation parameter, given in the title, and the corresponding modulation scale is plotted with a vertical black line. Colored solid curves represent the sensor gradient $S_x$, while dashed colored lines represent a linear fit in the regime interior to the modulation scale. Bluer curves correspond to larger $r_0$, and smaller phase aberrations, while redder curves are smaller $r_0$, with larger phase aberrations.}
    \label{fig:slope_test}
\end{figure}

To demonstrate the effect of optical gain in pedagogical scenario, we investigate the sensor response to the simplest aberration that can be present, tip and tilt. Using our simulation framework described earlier, additional tip and tilt can be injected on top of atmospheric phase aberrations, and the resulting sensor gradients can be measured. Because the sensor gradients are typically a function of pupil location for complex aberrations, this can be difficult to visualize, but for the case of tip and tilt, the sensor gradient can be reduced to a single number which is the average slope over the entire pupil. In Figure \ref{fig:slope_test} we show the average sensor gradient $S_x$ for various injected tilts, different values of $r_0$, and different modulation parameters.

Examining the response curves, it is clear than inside the modulation radius, the sensor response to tilt is appropriately linear. For the smallest modulation radius, the linearity extends even past this scale due to the presence of atmospheric aberrations causing the light to be further spread out over the pyramid face. This effect is less important at larger modulation radius. However, outside the modulation scale, the sensor begins to saturate, and the non-linearity onset causes the resulting measured gradient to asymptotically approach the maximum value of $S_x = 1$ due to normalization. 

\begin{figure}[h!]
    \centering
    \includegraphics[width=.8\textwidth]{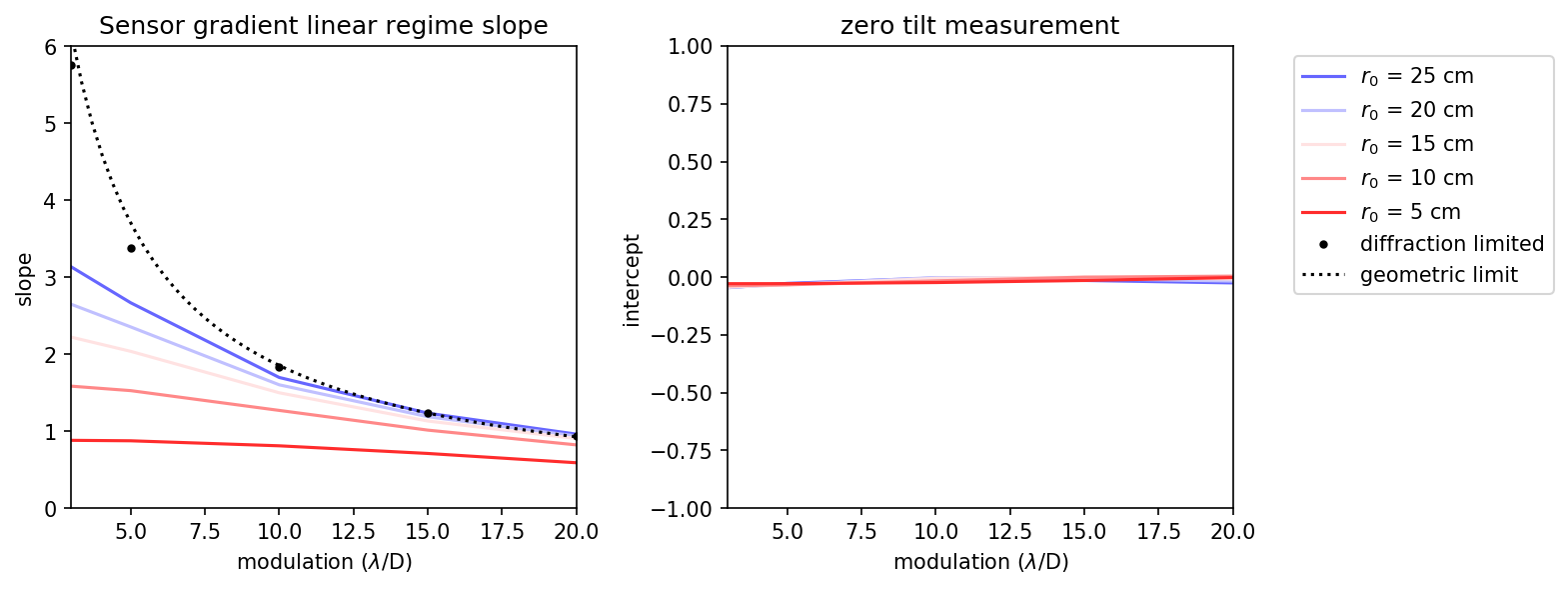}
    \caption{Summary of linear fits from Figure \ref{fig:slope_test}, as well as a comparison to a diffraction-limited case with zero phase aberrations and the analytic curve using the geometric approximation \cite{VERINAUD2004}.}
    \label{fig:slope_intercept}
\end{figure}

The previously discussed sensor response functions are fit with a linear model for the data points inside the modulation radius, and the slope and intercept of that line are summarized in Figure \ref{fig:slope_intercept}. The intercepts are generally close to zero, although not perfectly, due to the small number of atmospheric realizations that have been averaged over to generate the curves. Larger number of realizations could be averaged over to drive the intercepts closer to zero, but this is computationally prohibitive. However, the slope of the sensor response curve is much more interesting. The slope can be though of as a multiplicative factor which is needed to relate the input tilt to the output sensor gradient, and is thus directly related to the optical gain for tilt. Specifically, the optical gain should be $1/$slope.

A geometric analysis of the Pyramid Wavefront sensor \cite{VERINAUD2004} demonstrates that the relationship between sensor gradient and wavefront tilt should be inversely proportional to the modulation radius. Specifically, that
\begin{equation}
S_x = \frac{\lambda}{\alpha_\textrm{mod}\pi^2} \frac{\rmd \phi}{\rmd x}.
\end{equation}
However, we can see from the curves of slope vs modulation that this is not strictly true for all $r_0$. As $r_0 \rightarrow \infty$ becomes large, it approaches the inverse relationship between slope $\propto 1/\alpha_\textrm{mod}$ and modulation scale, but there is always a small linearization induced by the size of the diffraction-limited spot. Being able to recover the behavior of the analytic geometric approximation when $r_0$ is large is an interesting exercise, but the change in behavior is noticeably distinct when $r_0$ is small. 

The presence of large phase aberrations causes the light to be spread out over the pyramid face, reducing the measured sensor gradients, resulting in smaller measured slopes for a given tilt. This in turn requires a larger optical gain to compensate. The takeaway here is that the sensor response depends on the aberrations themselves, and to properly calibrate the sensor response to any aberration, the aberrations must already be known, and so the calibration problem is stuck in a loop, as one would hope to measure the aberrations with the sensor. This conundrum leads others to explore methods of instantaneously estimating the optical gain from the measurements, such a temporally dithering a low amplitude perturbation of a low order mode in a framework of optical gain tracking.\cite{esposito2020}

This simple example with tilt should clearly demonstrate the difficulty in calibrating the non-linear response regime of the Pyramid Wavefront sensor, but it would also be interesting to investigate this behavior for higher order modes. Due to the difficulty in plotting the sensor response function for complicated modes which are a function of pupil location as well as amplitude of the input perturbation, we do not directly investigate the sensor response curves for higher order modes. While, it may be possible to measure small modal perturbations on top of aberrated PSFs as was done previously for tip and tilt, this is complicated by the atmospheric aberration having non-zero projection onto the mode used as a perturbation.

\begin{figure}[h!]
    \centering
    \includegraphics[width=.7\textwidth]{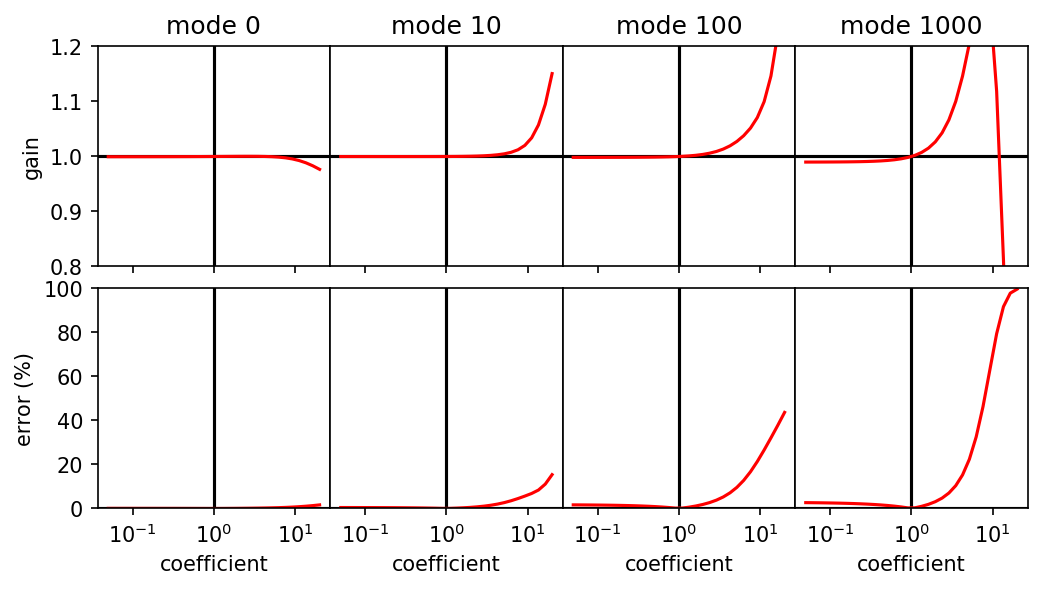}
    \caption{Modal optical gain calibration. Each column corresponds to a different modal perturbation on the DM. The first row is the measured optical gain for that mode and amplitude, and the bottom row is the RMS reconstruction error after the gain is applied.}
    \label{fig:modal_gain}
\end{figure}

To circumvent these issues, we instead investigate the reconstructed DM amplitudes for various modal perturbations of different amplitudes, with the atmosphere turned off. This allows us to see the non-linearity onset at a particular modal coefficient for each distinct mode by plotting the projection of the reconstructed DM onto the input vector. The results are plotted in Figure \ref{fig:modal_gain}. The optical gain reported is the scalar projection of the reconstructed DM vector onto the DM vector used as a perturbation, specifically
\begin{equation}
    \textrm{optical gain} = \frac{\langle \phi_\textrm{DM}, \phi_\textrm{recon} \rangle}{\langle \phi_\textrm{recon}, \phi_\textrm{recon} \rangle}.
\end{equation}

For all modes, perturbations with modal coefficients $\mathbf{v}_\textrm{DM} < 1$ result in measured optical gains $\sim 1$, and very small corresponding errors. However, with $\mathbf{v}_\textrm{DM} \gtrsim 1$, the non-linearity onset becomes apparent as the reconstructed DM no longer reconstructs the appropriate magnitude of the perturbation, and the resulting reconstruction error begins to increase. However, it is interesting to note that the particular value of the modal coefficient where this onset occurs depends on the actual mode itself. For low order modes the onset occurs later, at larger coefficients, while for high order modes the onset occurs sooner at low modal coefficients. For reference, a modal coefficient of $1$ corresponds to roughly $160 \pm 12$ nm of phase peak-to-valley for any particular mode. 

In addition, the fact that the reconstruction error continues to increase as the optical gain deviates from $1$ indicates that additional failure in the reconstruction of the shape is occurring. If instead the reconstructor produced the proper shape but rescaled by a constant, the reconstruction error would be small after the optical gain correction. The major takeaway here being that in order to properly calibrate the sensor for the non-linear regime, complete knowledge of the aberrations is needed, but that the interaction matrix framework is self-calibrating for small amplitude perturbations, which guarantee the success of the AO system during closed-loop operation.

\subsection{Simulation Results and Performance}

To mimic the real time system operation in our simulation it is necessary to approximate the real time closed loop behavior in a computationally feasible manner. For GPI 1, actual system delays from the end of the integration step, including CCD read, computation, and applying the resulting commands, can range from 1.2 frames to 1.6 frames at 1 kHz. \cite{Poyneer16} GPI 2 will include a faster computer and camera which could reduce these delays to 430 $\mu s$ or .43 frames at 1 kHz. In order to avoid the high temporal resolution sampling needed to resolve the effects of sub-frame delays, we restrict our approach to integer frame delays at 1 kHz. This allows us to efficiently mimic the real time system, while allowing computationally feasible calculation of the atmosphere with intervals of 1 ms timesteps. 

\begin{figure}[h!]
    \centering
    \includegraphics[width=.85\textwidth]{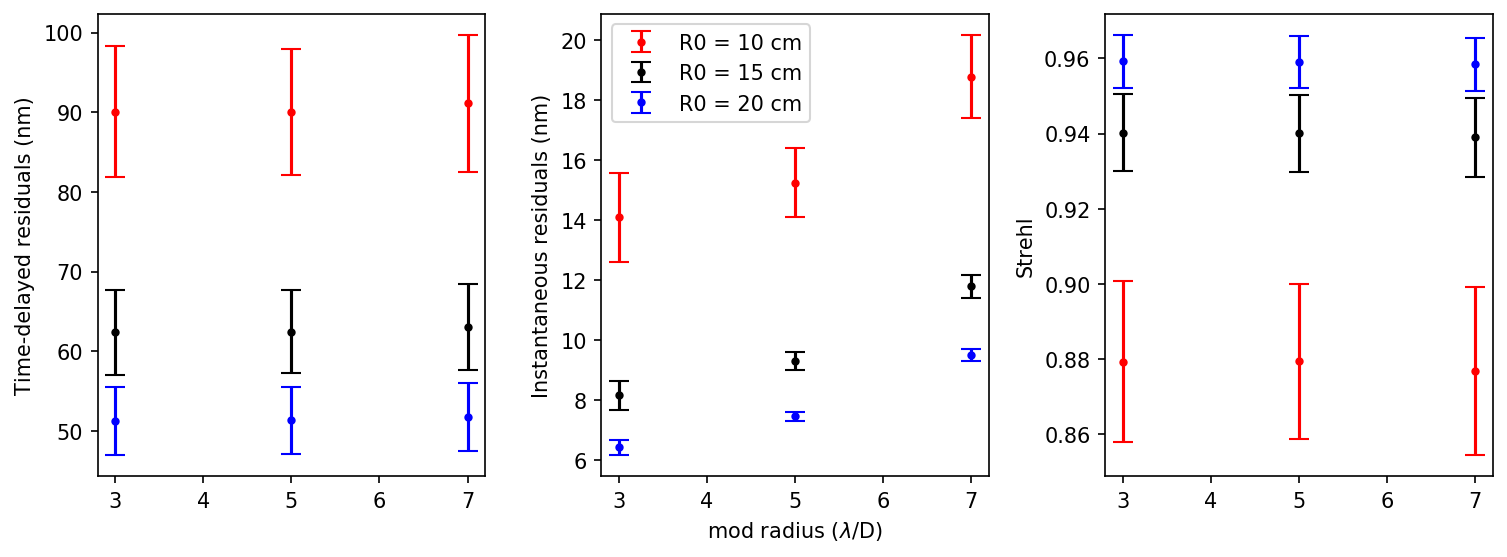}
    \caption{Performance metrics for an ideal AO system with the modal vector space interaction matrix. The left plot is the residual phase for the zero frame delay, which includes temporal errors due to the integration time. The center plot is the residual phase for an impossible -1 frame delay situation, as if the reconstruction was applied to the same atmosphere for which is it was measured, and represents the reconstruction error on a static aberration. In this plot it is clear that smaller modulations have correspondingly better reconstruction performance, but that this difference is small compared to the effect of the time delay error adding in quadrature. The right plot is the Strehl ratio of the resulting PSFs, which when measured in the image plane do not account for tip-tilt errors, as it just compares the maximum intensity to the ideal PSF.}
    \label{fig:peformance}
\end{figure}

If the delay is zero frames at 1 kHz, this approach essentially takes a measurement from the current system state, reconstructs the new DM coefficients, and applies those coefficients during the next timestep. This is an optimistic assumption that neglects additional temporal error due to computation time, which could be accounted for with a 1 frame delay. This more pessimistic scenario would comparatively overestimate the time lag error, and be more difficult to implement, requiring a memory buffer of some kind instead of a simple for loop. We report results from the zero frame delay, because it still accounts for temporal error due to integration time and is simpler to implement. Performance metrics are plotted in Figure \ref{fig:peformance} for various $r_0$ and modulation angle. These simulations use a maximal loop gain of 1 on an infinitely bright star with no WFS noise, while controlling the first 2000 modes in the DM vector space. After allowing the loop to converge for 5 timesteps, the simulation is run for an additional 100 timesteps to estimate the mean and variance of the metrics. 

\begin{figure}[h!]
    \centering
    \includegraphics[width=.5\textwidth]{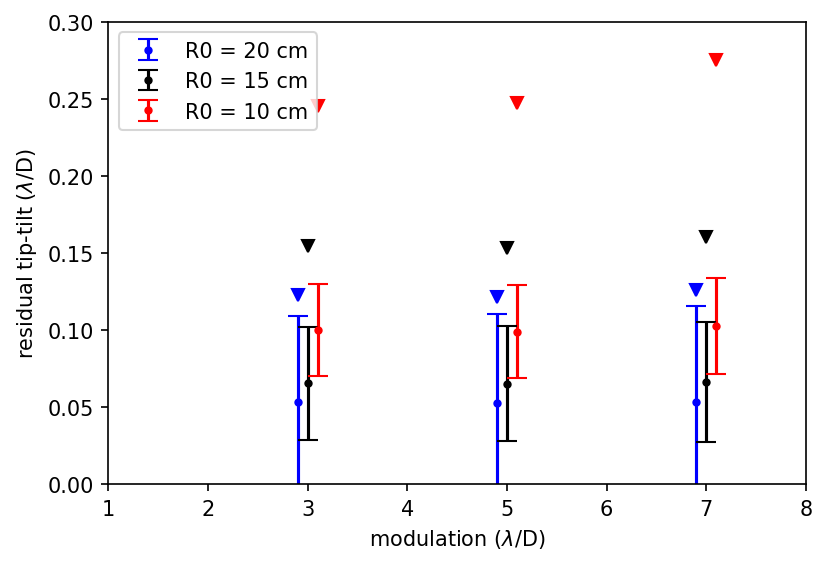}
    \caption{Residual tip tilt errors from the simulation. The error bars represent the mean and variance of the residuals, while the triangle represents the maximal value over the 100 timesteps.}
    \label{fig:tip_tilt}
\end{figure}

In addition to the performance metrics above, it is also interesting to investigate the residual tip and tilt by fitting a slope to residual phase. Image processing algorithms often struggle directly outside the edge of the coronagraphic mask \cite{nielsen2019}, and residual tip tilt errors make cause significant star light leakage through the coronagraph. \cite{lloyd2005} These results are plotted in Figure \ref{fig:tip_tilt}.

In each of these metrics, it is not obvious that the modulation parameter has much effect on the performance of the simulation, but this is only due to the idealization that the star is infinitely bright, and that there is no corresponding noise in the WFS. Further testing including WFS noise indicates that larger modulation scales suffer from WFS noise more strongly, as the light is spread more out between the four reimaged pupils more, and the corresponding measured gradients are smaller, needing a larger gain to reconstruct the input phase. In essence, larger modulation requires amplifying the effect of WFS noise.

Running the simulation with different WFS noise levels corresponding to different guide star brightness indicate that the simulation effectiveness breaks down when the average number of photons per subaperture is $\lesssim 1$. The noise on the computation of the gradients is order $\sim 100 \%$ and unstable modes develop. In these faint star situations it may be necessary to use a more robust framework of estimating the reconstructor, such as including priors on the noise covariance of the reconstruction \cite{vanDam2004}, or by running the system at a slower framerate such as 500 kHz to trade off temporal and reconstruction error. This kind of temporal averaging leverages the central limit theorem to cause the measurement errors to become more Gaussian, so that the linear inverse is a better approximation. A similar statement could be made regarding spatial averaging of the gradients among the sub-apertures.

\subsection{NCPA correction}

Non-common path aberrations (NCPA) are a well-known difficulty for a real system using a Pyramid Wavefront Sensor. \cite{esposito2020,chambouleyron2020} These aberrations are usually static aberrations in the science path of the instrument, which the wavefront sensor cannot see and correct. These aberrations degrade the final image quality if they remain uncorrected, but if they are known can be fixed by the deformable mirror. In this section we investigate the ability of our idealized AO system to correct NCPA in two distinct tests.

The problem can be succintly summarized as follows. Some aberration exists $\phi_\textrm{NCPA}$ which is observed in the final PSF, which is not seen from by the wavefront sensor. In order to mitigate this aberration, the AO loop is configured to drive the residual phase $\phi_\textrm{atm} + \phi_\textrm{DM} = -\phi_\textrm{NCPA} \neq 0$ not to zero but instead the inverse of the NCPA. This way the light reaching the science path is "pre-corrected" for the known static aberrations, and will result in a flat wavefront just before the final PSF is generated. This can be achieved generally by changing the reference zero state of the WFS $I_\textrm{WFS,0}$ to a new reference measurement set, with the DM phase equal to the negative of the NCPA phase.

\begin{figure}[h!]
    \centering
    \includegraphics[width=.85\textwidth]{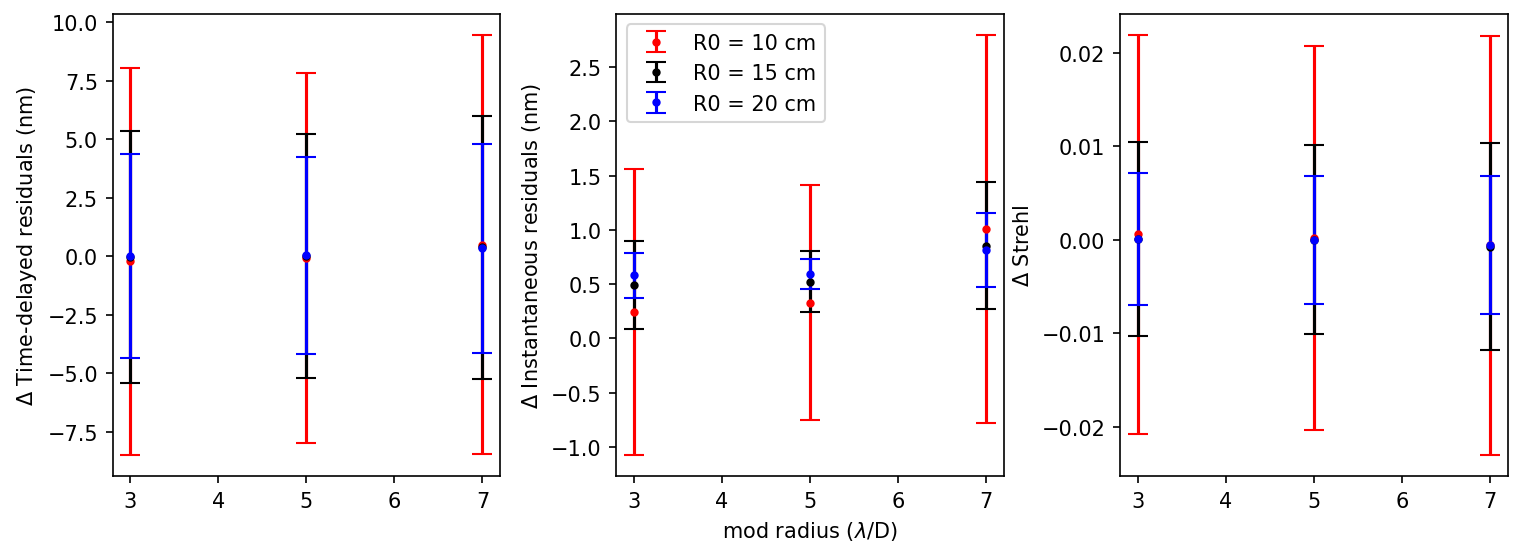}
    \caption{Change in performance metrics for the simulations described in Figure \ref{fig:peformance} with the addition of 50 nm RMS defocus NCPA. The small defocus is within the linear range of the sensor, and the performance is largely unaffected.}
    \label{fig:ncpa_performance}
\end{figure}

The first NCPA test is for a small defocus term, with 50 nm RMS phase, and is plotted in Figure \ref{fig:ncpa_performance}. The defocus term is specifically the Zernike polynomial \cite{ZERNIKE1934} $Z^0_2 \propto 2\sqrt{x^2+y^2}/R_\textrm{tel}-1$, which has been rescaled to the appropriate normalization. Small defocus terms could arise in the instrument path from slight axial disturbance of focusing optics, due to thermal flexure, changes in gravity, or mechanical tolerances in optical mounts. GPI 1 is known to suffer from roughly 50 nm RMS of defocus in the final image, and so this test is to investigate the severity of the impact of this known error. It remains an open issue to measure NCPA for the new and instrument, as changes to the optomechanics of the instrument may introduce new errors.

\begin{figure}[h!]
    \centering
    \includegraphics[width=.5\textwidth]{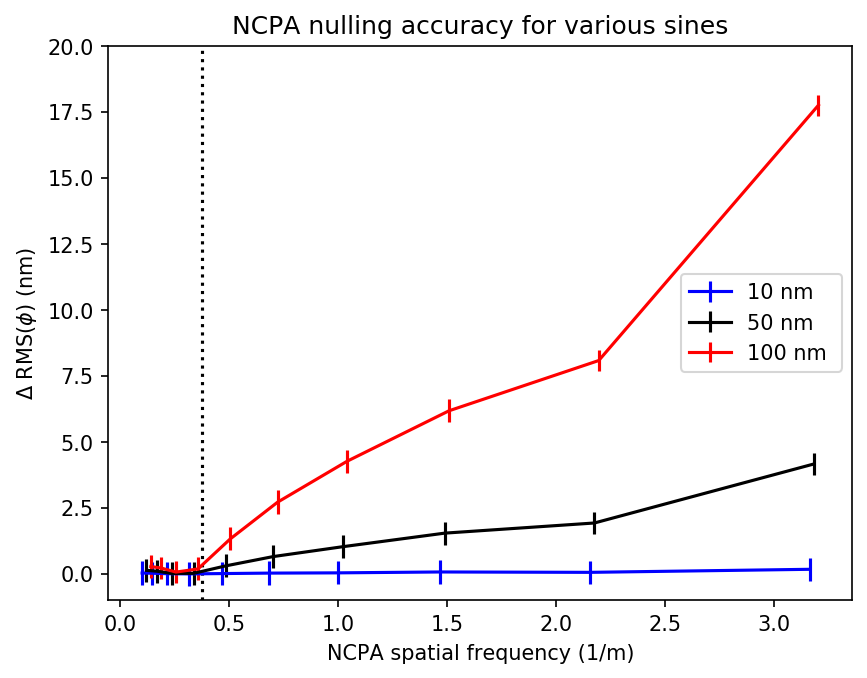}
    \caption{Change in residual phase for sine wave NCPA's of various spatial frequencies and amplitudes. For small amplitude perturbations, any spatial frequency is in the linear range, but large amplitude perturbation may introduce errors up to $\sim20\%$ for the highest spatial frequencies. The black dashed line represent the modulation scale. It appears that all of spatial frequencies inside the modulation scale are corrected regardless of amplitude, but this doesn't hold when investigating different modulation scales and may be a coincidence.}
    \label{fig:ncpa_sines}
\end{figure}

The second NCPA test investigates sine waves of different spatial frequencies and amplitudes, and is plotted in Figure \ref{fig:ncpa_sines}. 

Looking at the results of these tests, it appears that correcting NCPA is not an issue as long as the aberrations are small and inside the linear regime of the sensor. If the NCPA are large and would cause the sensor to be affected by non-linearity or saturation, then a deeper understanding is required. In particular, the coupling of the NCPA correction to the optical gain calibration is well known. \cite{esposito2020}

\section{GPI2 Performance Estimates}

The simulations described in the previous section could potentially be used to estimate the instrument performance, but they do not quite capture all of the complex behavior of the instrument and are very idealized. An additional software module for modelling the coronagraph would be necessary, and accurate coronagraph modeling would increase the computational difficulty significantly. \cite{Douglas2018} Typical observing sequences for high contrast imaging cover timescales of hours, while our simulations require days to compute only a few seconds of observing time. Since contrast is often limited by speckles, which can evolve due to changing wind and atmospheric conditions, recovering the proper distribution of light in the image plane is quite challenging. Furthermore, many sources of error exist in the instrument which is not properly accounted for in the simulation, including DM fitting error due to actuator influence functions and hysteresis, temporal errors due to sub-frame delays, proper modeling of the system control loop and gain optimization, and other systematics.

In order to avoid all of these top-down difficulties in understanding the instrument performance, it is rather simpler to just examine the actual instrument performance itself, and estimate improvements based on known fundamental changes. This section attempts that analysis. GPI 1 performance is well known \cite{nielsen2019} from the survey on Gemini South, and by comparing the atmospheric properties of Cerro Pachon and Mauna Kea, as well as changes to the system ETF, we estimate the improvement of the performance under a set of conservative but simplifying assumptions.

\subsection{The Error Transfer Function and the Modal Gain Optimizer}

\begin{figure} [h!]
\centering
\tikzstyle{block} = [draw, rectangle, 
    minimum height=2em, minimum width=4em]
\tikzstyle{sum} = [draw, circle, node distance=2cm]
\tikzstyle{input} = [coordinate]
\tikzstyle{output} = [coordinate]
\tikzstyle{pinstyle} = [pin edge={to-,thin,black}]
\begin{tikzpicture}[auto, node distance=3cm,>=latex']
    \node [input, name=input] {};
    \node [sum, right of=input] (sum) {};
    \node [block, right of=sum, pin={[pinstyle]above:$v(t)$},
            node distance=3.5cm] (WFS) {$W(s)$};
    \node [block, right of=WFS, node distance=4cm] (Read) {$A/D$};
    \node [output, right of=Read] (output) {y(t)};
    \node [block, below of=Read, node distance=1.5cm] (Ctrl) {$C(z)$};
    \node [block, left of=Ctrl] (D2A) {$D/A$};
    \node [block, left of=D2A] (Comp) {$e^{-\tau_{c}s}$};
    
    \draw [draw,->] (input) -- node[pos=0.99] {$+$} 
            node [] {$\phi_\textrm{atm}(t)$} (sum);
    \draw [->] (sum) -- node {$\phi_\textrm{res}$} (WFS);
    \draw [->] (WFS) -- node {} (Read);
    \draw [->] (Read) -- node[name=phi_c, near end] {$\phi_\textrm{meas}(t)$} (output);
    \draw [->] (phi_c) |- (Ctrl);
    \draw [->] (Ctrl) -- node {} (D2A);
    \draw [->] (D2A) -- node {} (Comp);
    \draw [->] (Comp) -| node[pos=0.95] {$-$} node [near end] {$\phi_\textrm{DM}$} (sum);
\end{tikzpicture}
\caption{Adaptive optics system control block diagram}
\label{fig:control}
\end{figure}
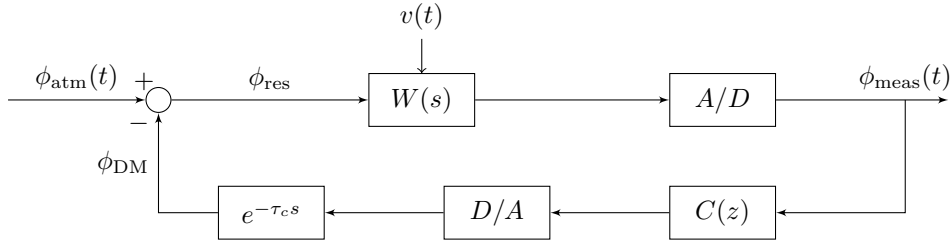

We model the AO system control using standard control theory techniques \cite{Poyneer16,roddier1999}. Figure \ref{fig:control} gives the block diagram of our hybrid continuous-discrete AO control system in the Laplace domain, where $s = i\omega$ and $\omega$ denotes the temporal frequency. The input is the turbulent wavefront $\phi_\textrm{atm}(t)$ and is continuously compensated by the combined woofer and tweeter phase $\phi_\textrm{DM}(t)$. The WFS measures the residual phase $\phi_\textrm{res}(t)$ with noise $v(t)$ and the WFS behavior can be characterized as an integration during one sampling period T
\begin{equation}
    W(s) = \frac{1-e^{-Ts}}{Ts}.
\end{equation}
The real time computer records the WFS signals and is modeled as a pure time delay $e^{-\tau_{r}s}$ due to the read-out time $\tau_{r}$. This process is shown as the A/D block and it outputs the discrete phase residual measurement $\phi_\textrm{meas}(t)$. The measurement is then sent into the discrete integral controller $C(z)$ which can be modelled as
\begin{equation}
    C(z) = \frac{g}{1-cz^{-1}},
\end{equation}
where $z=e^{sT}$ and the integrator constant is $c=0.999$. The matched pole-zero method \cite{franklin} is used to convert between the continuous form $C(s)$ and the discrete form $C(z)$. The control gain $g$ is unique to each mode and is optimized based on stability margin analyses. The controller outputs discrete control voltages of the tip-tilt stages and the DM and the D/A converter hold the voltages constant during each sampling period T, so it can be modeled as a zero-order hold
\begin{equation}
    D(s) = \frac{1-e^{-Ts}}{Ts}.
\end{equation}
Note that though $W(s)$ and $D(s)$ have the same mathematical form, they have different physical meanings. The computational time delay is modeled as $e^{-\tau_{c}s}$ and its block is placed after the D/A converter even though it represents the total computational delay $\tau_{c}$ throughout one control loop. The error transfer function $ETF(s) = {\phi_\textrm{res}(s)}/{\phi_\textrm{atm}(s)}$ is the ratio between the residual phase and the uncompensated wavefront, so it characterizes the AO system performance at different temporal frequencies. The ETF can be evaluated as
\begin{equation}
    \textrm{ETF}(s) = \frac{1}{1 + L(s)},
\end{equation}
where $L(s) = W(s)D(s)C(e^{s})e^{-\tau s}$ is the open loop transfer function and $\tau=\tau_{r}+\tau_{c}$ is the total delay from read-out and compute time. Simplifying the open loop transfer function results in
\begin{eqnarray}
    |L(s)| =& \frac{2g(1 - \cos (\omega T))}{(\omega T)^2 \sqrt{1 - 2c \cos (\omega T) + c^2}}, \\
    \textrm{arg}(L(s)) =& \textrm{arctan}\left(\frac{-c\sin(\omega T)}{1-c\cos(\omega T)}\right) + 2 \textrm{arctan}\left(\frac{\cos(\omega T)-1}{\sin(\omega T)}\right) - 2\pi \omega \tau.
\end{eqnarray}

\begin{figure}[h!]
    \centering
    \includegraphics[width=.4\textwidth]{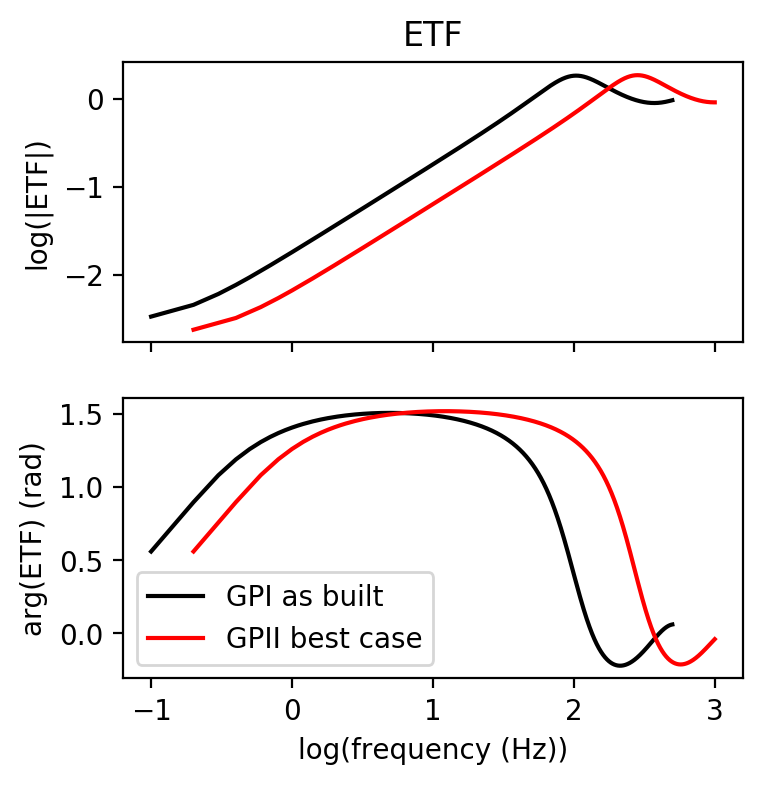}
    \caption{Bode plot for GPI 1 and 2 Error Transfer Functions.}
    \label{fig:etf}
\end{figure}

Our modal gain optimizer algorithm finds the optimal control gain g for each mode while ensuring the system's robustness to modelling errors and noise. We formulate a constrained optimization problem which finds the maximal gain for each possible value of the time delay $\tau$ while maintaining an open loop gain $L(s)>$ 2.5 and a phase margin $\textrm{arg}(L(s))>$ 45$^{\circ}$. \cite{friedland} We report the following values for the optimal gains in Table \ref{tab:gain}. These margins are chosen empirically and they ensure the system's stability under disturbances and uncertainties. The whole pipeline will be useful later when we analyze and tune the response of GPI 2 under various delays. Figure \ref{fig:etf} compares the bode plot of the best case ETF in GPI 2 to current GPI. We can see the bandwidth improves from 50 Hz to 139 Hz and the rejection of slower signals improves by a factor of 8 in terms of power, which equates to mean-squared error and speckle intensity.

\begin{table}[h!]
    \centering
        \caption{Optimal loop gains at various delays}
    \label{tab:gain}
    \begin{tabular}{c | c c c c c c c c c c}
    $\tau$ (ms) & 0.800 & 0.758 & 0.717 & 0.676 & 0.635 & 0.594 & 0.553 & 0.512 & 0.471 & 0.430 \\
    \hline
    g & 0.306 & 0.319 & 0.333 & 0.349 & 0.367 & 0.385 & 0.407 & 0.431 & 0.457 & 0.488
    \end{tabular}
\end{table}

\subsection{Atmospheric Comparison}
\begin{figure}[h!]
    \centering
    \includegraphics[width=.6\textwidth]{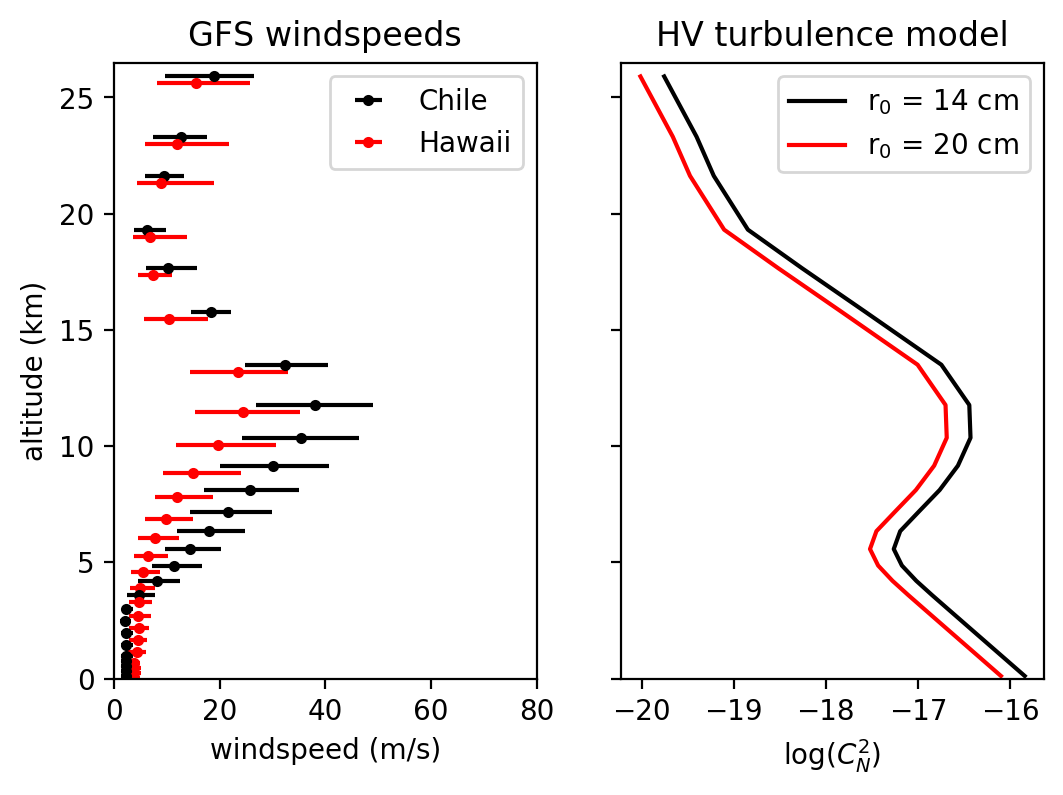}
    \caption{The atmosphere model used for our analysis including windspeeds from the NOAA GFS \cite{gfs} for the two sites and a heuristic turbulence profile. The error bars demonstrate the 25th and 75th percentile wind velocity while the center point is the median.}
    \label{fig:atmosphere}
\end{figure}

In order to evaluate the instrument response, a model of the atmosphere for each of the two sites, Cerro Pachon, Chile, and Mauna Kea, Hawaii, will be needed to compare the relative difficulty of observing in each environment. To do this, we sample a large quantity of wind velocities from the NOAA Global Forecast System (GFS) \cite{gfs} to build a representative distribution. Furthermore, we assume the atmospheric turbulence has a Kolmogorov power spectrum
\begin{equation}
    |\Phi(k)|^2 \propto C_N^2 k^{-11/3}
\end{equation}
and that the structure function $C_N^2$ varies with height according to the Hufnagel-Valley model \cite{Canuet2015AtmosphericTP}, given by
\begin{equation}
    C_N^2 = A \Big[ 2.2\times 10^{53} h^{10}\Big(\frac{v}{27}\Big)^2 e^{-h/1000} + 1\times 10^{-16} e^{-h/1500.} \Big]
\end{equation}
where $A = e^{\mathcal{N}(0,1)}$ is a random number for each layer and time instance, but the average can be computed using $\langle A \rangle \sim 2.7 $. It is then straightforward to compute the Fried Parameter \cite{hardy1998adaptive}, using 
\begin{equation}
    r_0 = \Big[ .423 k^2 \int  C_N^2 \rmd h \Big]^{-3/5}
\end{equation}
so we can rescale the structure constant to result in any $r_0$ we wish. Using nominal values of $r_0 = 14$ cm and $20$ cm for Chile and Hawaii, respectively, based on seeing measurements from site analyses, \cite{racine1991,Tokovinin2006} the resulting profiles appear in Figure \ref{fig:atmosphere}. While the $C_N^2$ model may not be truly identical between the two sites, obtaining consistent empirical data for the two mountains is a challenge and slight deviations from the general behavior would only marginally modify the results compared to changes in $r_0$.

\subsection{Estimating Contrast Enhancement}

According the Perrin et al 2003 \cite{perrin}, any PSF can be expanded in a taylor series
\begin{eqnarray}
\textrm{PSF} &\approx& \textrm{PSF}_0 + \textrm{PSF}_1 + \textrm{PSF}_{2,\textrm{halo}} + \textrm{PSF}_{2,\textrm{strehl}},
\end{eqnarray}
\begin{eqnarray}
\textrm{PSF}_0 &=& aa^* \\
\textrm{PSF}_1 &=& 2\textrm{Im}[a(a^* \star \Phi^*)] \\
\textrm{PSF}_{2,\textrm{halo}} &=& (a \star \Phi)(a^* \star \Phi^*) \\
\textrm{PSF}_{2,\textrm{strehl}} &=& -\frac{1}{2}[a(a^* \star \Phi^* \star \Phi^*) + a^*(a \star \Phi \star \Phi)] .
\end{eqnarray}
If we assume there are no amplitude errors, $a = \delta(0)$ implies $\textrm{PSF}_{2,\textrm{halo}} = |\Phi|^2$. Additionally, if $\phi$ is even, $\Phi$ is real, and $\textrm{PSF}_1 = 0$. Furthermore, we can ignore $\textrm{PSF}_{2,\textrm{strehl}}$ since $\Phi \star \Phi \leq |\Phi|^2$. Under all of these conditions, we arrive at our estimate of the scaling 
\begin{equation}
    \textrm{PSF} \propto |\Phi|^2.
\end{equation}
However, if we assume the final image contrast reached by post-processing algorithms is not set by the raw intensity but rather by the limits of the photon noise, which scales like the square root of the intensity, the contrast improvement will scale like $|\Phi|$ and not $|\Phi|^2$. This conservative approach essentially assumes that post-processing algorithms are already perfect, and may not always be true especially for very bright stars.

To evaluate the relative performance of the instruments on the two mountains, we perform a comparative analysis of the residual phase. At each image angular separation $\theta$, speckles present will be due to phase aberrations with mode lengths $p$ such that $\theta = \frac{\lambda}{p}$. We consider only H-band so $\lambda = 1.6$ $\mu$m, and separations in the range $\theta \in (0.156,1.711)$ arcsec, corresponding to phase aberrations of mode length $p \in (0.193, 2.116)$ meters. The wavevector $k = 2\pi/p$ for each mode.
\begin{figure}[h!]
    \centering
    \includegraphics[width=.6\textwidth]{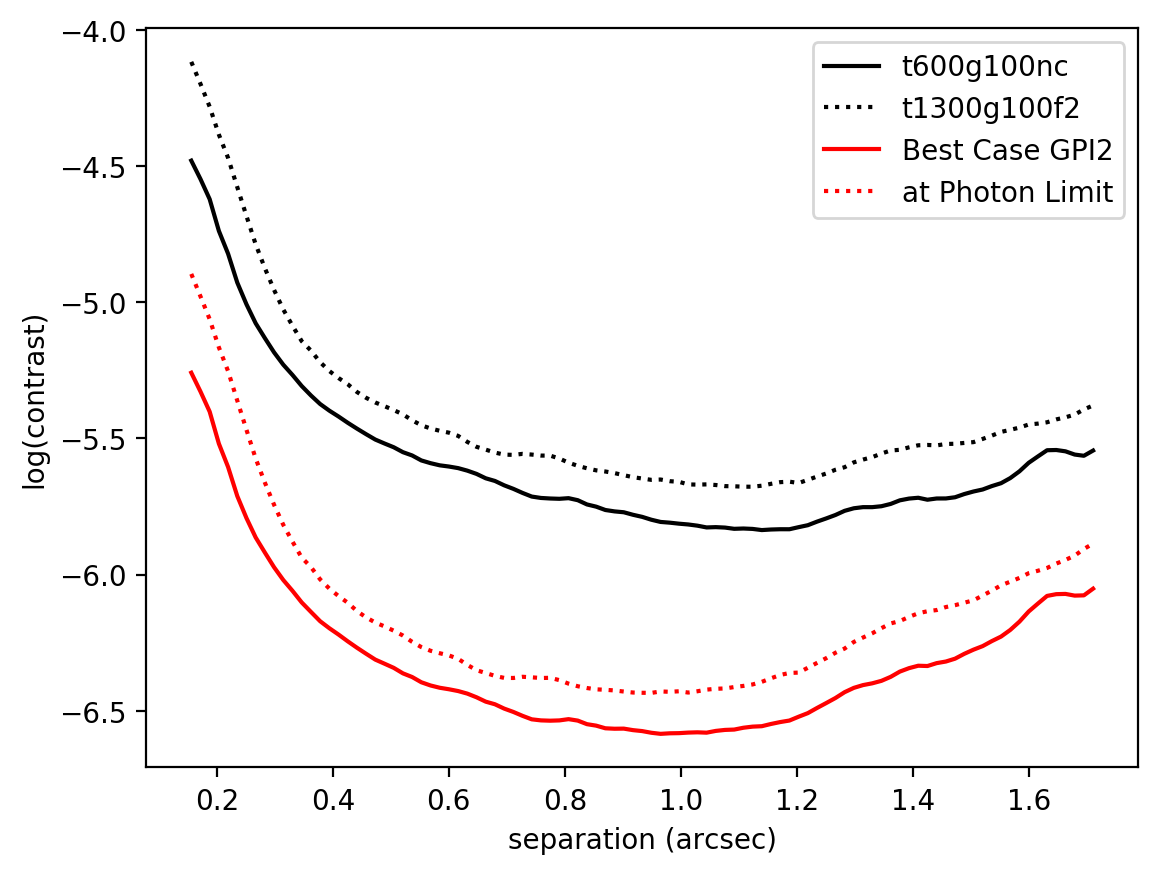}
    \caption{GPI2 contrast for median Mauna Kea atmospheric conditions estimated by rescaling GPI performance on Cerro Pachon using Equation (30). GPI's sensitivity is calculated by taking the median contrast curve of the GPIES survey after speckle subtraction and matched filtering according to the method described in Ruffio et al 2017. \cite{jb}}
    \label{fig:contrast}
\end{figure}

For each mode length $p$, we assume its wavevector is oriented parallel to the wind direction so that the temporal frequency of its oscillation is maximal, which is a worst-case scenario. Under this assumption, the temporal frequency of the oscillation is $f = v/p$, where $v$ is the wind velocity. If we also assume that the AO loops corrects each atmospheric layer independently, the magnitude of the ETF can be used to estimate the total residual phase after summing all of the contributions from the various layers. 
\begin{equation}
    \textrm{contrast enhancement} = \frac{\sum_{i}\sqrt{C_{N,\textrm{Hawaii}}^2(z_i)}\Delta z_i \times \Big|\textrm{ETF}_{\textrm{GPI2}}\Big(\cfrac{v_\textrm{Hawaii}(z_i)}{p}\Big) \Big|}{\sum_{i} \sqrt{C_{N,\textrm{Chile}}^2(z_i)}\Delta z_i \times \Big|\textrm{ETF}_{\textrm{GPI}}\Big(\cfrac{v_\textrm{Chile}(z_i)}{p}\Big)\Big|}
\end{equation}
Here, $i = 0,1,...,25$ is the index that runs over the layers at altitudes $z_i$ in the atmosphere model. Since the power spectrum $|\Phi|^2$ depends on the $C_N^2$ and $k$, with some constant pre-factor, those last terms factor out in the ratio and cancel. However, each layer contributes proportionally to its thickness $\Delta z_i$, which is not constant for our atmosphere model and therefore cannot be factored. This is because the accumulated phase from each layer is proportional to the optical path length (implicitly this assumes the various layers all have the same index of refraction, which isn't strictly true, as it weakly depends on pressure and temperature.)  The wind velocities $v$ are taken from the respective windspeed distributions on the two sites, either using a median or some other percentile. This metric estimates the relative final contrast between the two situations. An example is demonstrated in Figure \ref{fig:contrast}.

\begin{figure}[h!]
    \centering
    \includegraphics[width=.8\textwidth]{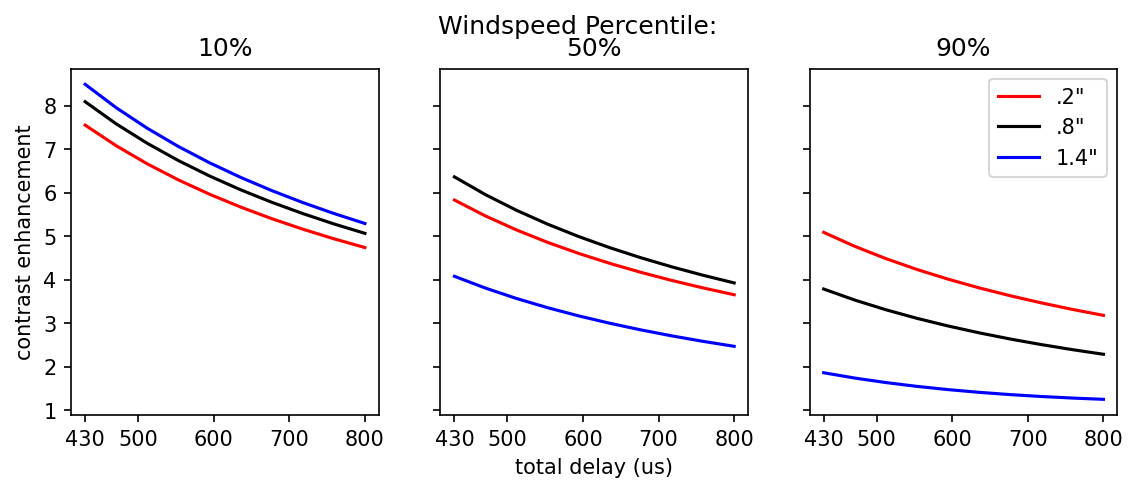}
    \caption{The contrast enhancement factor for various AO system delays during different weather conditions given by the percentile of windspeed and image separations. In general shorter delays are always better, but exactly how much depends on the image location and atmospheric conditions.}
    \label{fig:delays}
\end{figure}

For a final series of estimates, we use intermediary ETFs generated with different delays, whose gains are reported in Table \ref{tab:gain}, across a few different windspeed scenarios. The contrast enhancement is reported for three different image separations across different delays $\tau$ in Figure \ref{fig:delays}. In general it is clear that a faster computer is always better. However, it is interesting to note that for the best possible conditions Mauna Kea is much nicer than Cerro Pachon, although this does not consider variations in $r_0$ that may accompany the variations in windspeed. 

\section{Conclusions}

In conclusion, we have demonstrated the efficacy of the pyramid wavefront sensor as a replacement for the Shack-Hartmann sensor during the GPI 2 upgrade in simulation. Our simple reconstruction framework uses an efficient orthogonal basis derived from principal components of a model atmosphere, and computes the sensor linear response to small perturbations in an interaction matrix framework ideal for closed loop operation. The difficulties associated with the non-linearity of the pyramid response are discussed regarding optical gain calibration and non-common-path-aberration correction. However, measuring the NCPA remains an open problem. In the second half, our analysis pivots to a semi-empirical approach to estimate the final performance of the instrument, using a comparative analysis of the two system ETFs and the atmospheres corresponding to the two observatory sites. Our analysis shows that under the best observing conditions, final contrast at small separations could improve by nearly an order of magnitude, opening the door to observing fainter and closer in planets than before. A future survey on Gemini North could reveal new trends and resolve prior conflicts in giant planet demographics and formation mechanisms, especially in regard to hot vs. cold start planets.

\appendix
\newpage
\section{Atmospheric Model Calibration}

\begin{figure}[ht]
\centering
\begin{minipage}[b]{.4\linewidth}
\includegraphics[width=\textwidth]{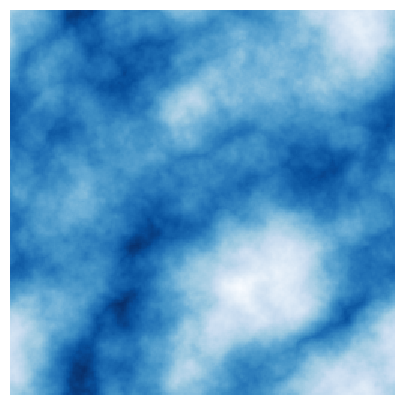}
    \caption{Kolmogorov phase aberrations at the bottom of the atmosphere.}
    \label{fig:kolmo}
\end{minipage}
\quad
\begin{minipage}[b]{.55\linewidth}
\includegraphics[width=.85\textwidth]{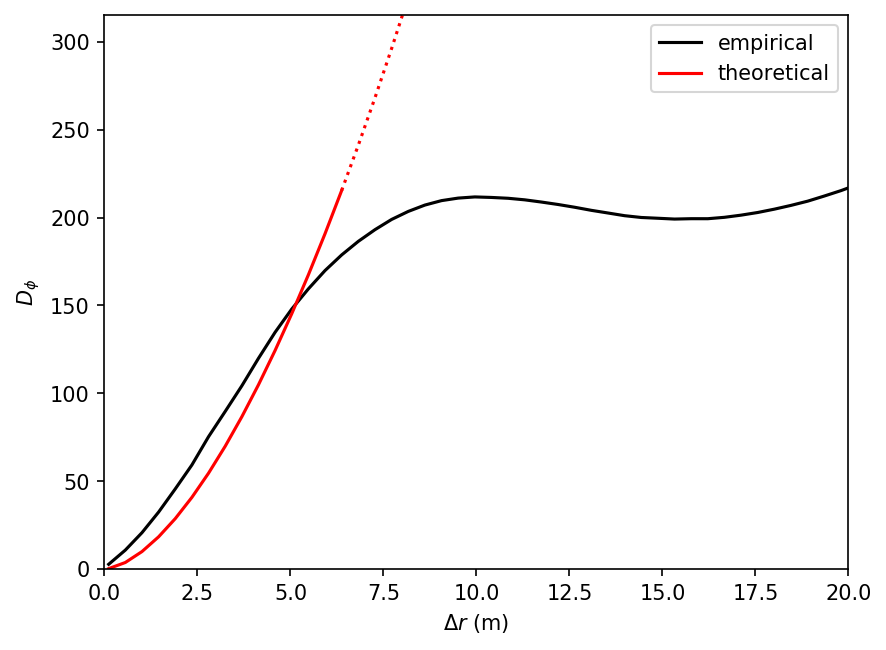}
    \caption{Phase Structure function for the simulation compared to the theoretical value. The empirical function has been rescaled by a constant which minimizes the $\chi^2$ in the region corresponding to the solid line. The dashed line is outside the fit region.}
    \label{fig:dphi}
\end{minipage}
\end{figure}

In order to evaluate the effectiveness of a pyramid WFS, we adopt the atmosphere model described in \cite{madurowicz2018} and later \cite{madurowicz2019}, which broadly includes Fresnel propagation of light through multiple turbulent Kolmogorov phase screens at various altitudes. Due to a limitation of the numerical implementation coercing complex number's phases into the range of $(-\pi, \pi)$, the resulting phase maps are "unwrapped" along discontinuities greater than $2\pi$, and rescaled in order to calibrate the Fried Parameter of the observation. An example of the resulting phase as seen at the bottom of the atmosphere is given in Figure \ref{fig:kolmo}.
Using the definition of the phase structure function given in Hardy \cite{Hardy1998}
\begin{equation}
    D_\phi(\Delta r) = \langle[\phi(r+\Delta r) - \phi(r)]^2\rangle
\end{equation}
we can compute the value of $D_\phi$ for the simulated atmosphere at various physical separations $\Delta r$. This is then comparable to the theoretical value derived later in Hardy,
\begin{equation}
    D_\phi(r) = 6.88 \Big(\frac{r}{r_0}\Big)^{5/3},
\end{equation}
which is demonstrated in Figure \ref{fig:dphi}

\begin{figure}[h!]
    \centering
    \includegraphics[width=\textwidth]{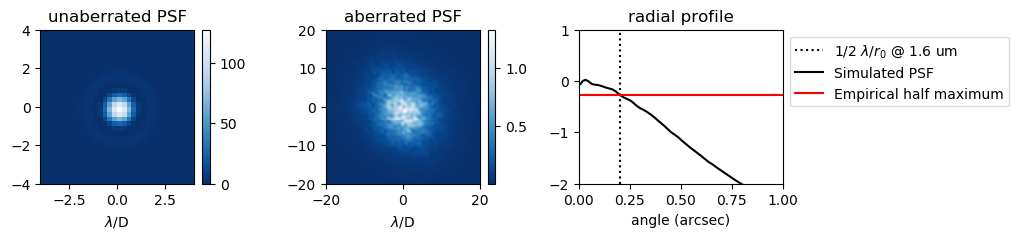}
    \caption{Comparison of an unaberrated PSF to the fully aberrated PSF. The colormap is proportional to the square root of the Intensity in the image plane. The radial profile of the aberrated PSF is given on the right, and its y-axis is proportional to the log of the intensity. As can be seen, the half maximum of the aberrated PSF occurs at the radial separation equivalent to 1/2 $\lambda/r_0$, indicating the calibration of the atmosphere is valid.}
    \label{fig:psftest}
\end{figure}

While the empirically derived structure function agrees well with the theory for small values of $\Delta r$, it begins to deviate when $\Delta r$ becomes large due to the finite box dimension being used the generate the Kolmogorov phase screens. Because the phase screens are generated by Fourier transforming a k-space power law with noise, periodic boundary conditions are enforced, and certain pairs of pixels are more correlated than they ought to be in a truly scale-invariant fractal. Due to this limitation, we choose to fit the empirical structure function to the theoretical for a range of $\Delta r \in (0, 6.3)$ meters. This arbitrary selection is chosen to coerce the FWHM of the aberrated PSF to the correct angular scale, demonstrated in Figure \ref{fig:psftest}.

\begin{figure}[h!]
    \centering
    \includegraphics[width=.5\textwidth]{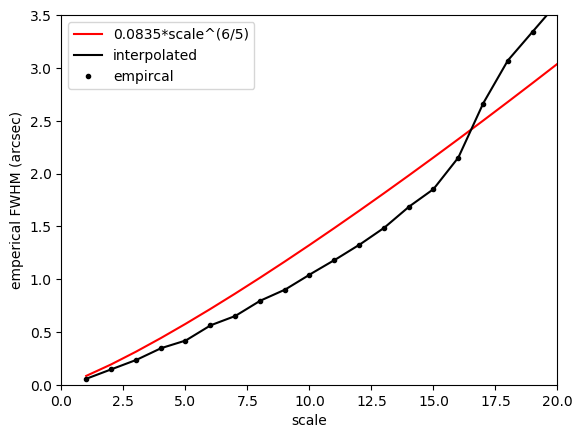}
    \caption{Comparing the simulated FWHM of the PSF to the value of the scaling constant applied to the phase aberrations.}
    \label{fig:fwhm_scale}
\end{figure}

However, this method of calibrating the phase is somewhat indirect, using the phase structure function as an intermediary calibrator to connect the aberrated PSF FWHM to a multiplicative rescaling factor of the phase aberrations. However, because we can simulate any phase rescaling factor, and measure its corresponding simulated FWHM, we can semi-empircally connect these two directly, which is demonstrated in Figure \ref{fig:fwhm_scale}. Combining the known theoretical scaling laws that the PSF FWHM $\propto \frac{\lambda}{r_0}$, and that the phase structure function $D_\phi \propto \Big(\frac{1}{r_0}\Big)^{5/3} \propto \phi^2$, we can conclude that the PSF FWHM $\propto \phi^{6/5}$. Both the simulated curve generated from one specific atmospheric realization and the theoretical best fit power law with index 6/5 are plotted in Figure \ref{fig:fwhm_scale}. While any particular realization may deviate from the theoretical mean scaling due to particular noise fluctuations, on average, multiple realizations share the same scaling behavior. This very empirical approach to calibrating allows us to find exactly the rescaling factor needed to achieve a particular PSF FWHM by interpolating known tested values.

\bibliography{report} 
\bibliographystyle{spiebib} 

\end{document}